\documentclass{article}
\usepackage[latin9]{inputenc}
\usepackage{geometry}
\usepackage{units}
\usepackage{amsmath}
\usepackage{amssymb}

\makeatletter

\providecommand{\tabularnewline}{\\}

\newcommand{\lyxaddress}[1]{
\par {\raggedright #1
\vspace{1.4em}
\noindent\par}
}

\usepackage{makeidx}
\usepackage{amsthm}\usepackage{epsfig}\usepackage{amsfonts}\usepackage{bbm}\usepackage{nicefrac}
\usepackage{units}

\makeatother

\begin{document}

\title{\textbf{\Large{}Eigenlogic: a Quantum View }\\
\textbf{\Large{} for Multiple-Valued and Fuzzy Systems }}

\author{François Dubois\textbf{$^{a,b}$} and Zeno Toffano\textbf{$^{c,d}$}}

\date{July 22, 2016}

\maketitle

\lyxaddress{}

\lyxaddress{\begin{center}
$^{a}$ LMSSC, Conservatoire National des Arts et Métiers, Paris,
France\\
 $^{b}$ Department of Mathematics, University Paris-Sud, Orsay, France\\
$^{c}$ Telecom Dept., CentraleSupélec, Gif-sur-Yvette, France\\
 $^{d}$ Laboratoire des Signaux et Systèmes, UMR8506-CNRS, Université
Paris-Saclay, France\\
\textit{francois.dubois@cnam.fr ,  zeno.toffano@centralesupelec.fr}\\

\par\end{center}}
\begin{abstract}
We propose a matrix model for two- and many-valued logic using families
of observables in Hilbert space, the eigenvalues give the truth values
of logical propositions where the atomic input proposition cases are
represented by the respective eigenvectors. For binary logic using
the truth values $\{0,1\}$ logical observables are pairwise commuting
projectors. For the truth values $\{+1,-1\}$ the operator system
is formally equivalent to that of a composite spin $\nicefrac{1}{2}$
system, the logical observables being isometries belonging to the
Pauli group. Also in this approach fuzzy logic arises naturally when
considering non-eigenvectors. The fuzzy membership function is obtained
by the quantum mean value of the logical projector observable and
turns out to be a probability measure in agreement with recent quantum
cognition models. The analogy of many-valued logic with quantum angular
momentum is then established. Logical observables for three-value
logic are formulated as functions of the $L_{z}$ observable of the
orbital angular momentum $\ell=1$. The representative $3$-valued
$2$-argument logical observables for the $\mathrm{Min}$ and $\mathrm{Max}$
connectives are explicitly obtained. \quad{}{{[}Contribution to
the conference ``Quantum Interaction 2016'', San Francisco USA,
20-22 July, 2016{]}} 
\\

\end{abstract}

\section{Introduction}

Quantum logic developed by Birkhoff and von Neumann in their seminal
article in 1936 \cite{BvN1936} considers logical propositions as
subspaces of a quantum state Hilbert space. As will be shown hereafter
and also underlined in \cite{Yi16}, these subspaces can be viewed
as eigenspaces of projectors, the projectors corresponding to logical
propositions. A true proposition is then associated to the eigenvalue~+1.
The representation of logical propositions in a vector space could
be of interest in modern semantic theories such as distributional
semantics, for example using the ``Hyperspace Analogue to Language''
algorithm as was done in \cite{BTMD14}, or in connectionist models
of cognition \cite{BB12}.

In this work we show that a proposition in a logical system can be
represented by an observable in Hilbert space. When interpreted in
the context of quantum mechanics this model uses finite dimensional
projectors and angular momentum observables. Conversely, a quantum
system when considered in its eigenspace is formally equivalent to
a logical propositional system. The view here, which comes under the
name of ``Eigenlogic'' (for the original motivation and more detailed
discussion see \cite{To15}), considers that the eigenvalues of the
logical observables are the truth values of a proposition and the
associated eigenvectors correspond to the different input atomic propositional
cases. When considering vectors outside of the eigensystem this view
leads to a ``fuzzy'' measure of the degree of truth of a logical
proposition.

\noindent In our model for binary valued logic, using numbers $\{0,\,1\}$,
the logical observables are pairwise commuting projectors. The model
is extended to the other binary system using numbers $\{+1,-1\}$,
differences reside in the symmetry of the corresponding logical observables.
In the latter case the observables are equivalent to quantum spin
$\nicefrac{1}{2}$ observables, no more idempotent projectors but
isometric self-inverse reflection observables squaring to $1$. These
are equivalent to the recently proposed ``quantum Boolean functions''
\cite{MO08} developed in the context of the research topic ``Fourier
analysis of Boolean functions'' having many applications in theoretical
computer science, information theory and also in social decision and
voting theory. We then propose an algebraic generalization, based
on the finite-elements method, that can be applied to whatever $m$-value
$n$-argument logical system.

\noindent The paper is organized as follows: we start with Boolean
two-valued $\{0,1\}$ logic and we demonstrate important expressions
for the projector observables in the $2$-argument case indicating
also the general method for $n$-arguments. The case for binary values
$\{+1,-1\}$ is then presented. Then we consider the case for fuzzy
logical propositions and give the method for calculating fuzzy membership
functions by using the Born rule and show that these functions can
be identified with probabilities. The last section is devoted to the
many-valued systems ($m>2$) the case of $3$-valued $2$-argument
logic is discussed with some examples of applications.

\section{Two-valued Eigenlogic}

\subsection{Projector two-valued logic}

We will consider a two-dimensional rank-$1$ projector $\boldsymbol{\Pi}$
acting on a single set. What are the expected outcomes when applying
this projector? If, for example, vector $|a>$ corresponds to an element
of the set, the following matrix equation will be verified: $\boldsymbol{\Pi}\cdot|a>=1\cdot|a>$.
The value $1$ being the eigenvalue of the projector associated with
the eigenvector $|a>$. Interpretable results \cite{To15} considered
in a two-value $\{0,1\}$ logical system will correspond to the possible
eigenvalues $0$ and $1$, where $0$ is the result for elements not
belonging to the set. So in this way a question concerning the proposition
of belonging or not to a particular set, will have as an answer one
of the two eigenvalues. The ``true'' value $1$ will correspond
to the eigenvector $|a>$, now named $|1>$, and the ``false'' value
$0$ will correspond to the complementary eigenvector $|\overline{a}>$,
named $|0>$. When these properties are expressed in matrix form:
vectors $|1>$ and $|0>$ become $2$ dimensional orthonormal column
vectors and the projection operators $2\times2$ square matrices.
This gives: 
\[
|1>\,\,=\,\,\left(\begin{array}{c}
0\\
1
\end{array}\right)\,,\qquad|0>\,\,=\,\,\left(\begin{array}{c}
1\\
0
\end{array}\right).
\]
The choice of the position of the value $1$ in the column vectors
is arbitrary, here it follows the quantum information convention for
a ``qubit-$1$'' \cite{NC00}. As usual in Quantum Mechanics we
can find the set of projectors that completely represent the quantum
system, in particular by lifting the eventual degeneracy of the eigenvalues.
Here eigenvalues are always equal to 0 or 1 and the question about
the multiplicity of eigenvalues is natural. In this contribution we
focus on different projective structures that completely define the
logical system. In the very simple case where 0 and 1 are both not
degenerate eigenvalues, the projectors relative to the eigenvector
basis take the form: 
\begin{equation}
\boldsymbol{\Pi}_{1}\,\,=\,\,\boldsymbol{\Pi}\,\,=\,\,\left(\begin{array}{cc}
0 & 0\\
0 & 1
\end{array}\right)\,,\qquad\quad\boldsymbol{\Pi}_{0}\,\,=\,\,\boldsymbol{I}\,-\,\boldsymbol{\Pi}\,\,=\,\,\left(\begin{array}{cc}
1 & 0\\
0 & 0
\end{array}\right)\,.\label{eq:projPi}
\end{equation}
We systematically consider all the possible structures of such projectors.
When representing logic with $n$ atomic propositions using projectors
various possibilities are intrinsically present in a unique structure
with $\,2^{2^{n}}\,$ different projectors. Once the eigenbasis is
chosen the remaining structure is intrinsic.

For example the two projectors shown in equation (\ref{eq:projPi})
are complementary and idempotent. One can give a general expression
of a one-argument ``logical observable'' as an expansion over the
commuting projectors $\boldsymbol{\Pi}_{0}$ and $\boldsymbol{\Pi}_{1}$
spanning the vector space: 
\begin{equation}
\boldsymbol{F}\,\,=\,\,f(0)\:\boldsymbol{\Pi}_{0}+f(1)\:\boldsymbol{\Pi}_{1}\,\,=\,\,\begin{pmatrix}f(0) & 0\\
0 & f(1)
\end{pmatrix}\label{eq:Operator-Dec1}
\end{equation}
the coefficients $f(0)$ and $f(1)$ in the expansion are the truth
values of the corresponding $\{0,1\}$ Boolean logical connective.
Equation (\ref{eq:Operator-Dec1}) represents the spectral decomposition
of the operator and because the eigenvalues are real the logical operator
is Hermitian and can thus be considered as a quantum observable. In
this way, in Eigenlogic, the truth values of the logical proposition
are the eigenvalues of the logical observable. One can then construct
the $4$ logical observables corresponding to the $4$ one-argument
Boolean connectives: $\boldsymbol{A}=\boldsymbol{\Pi}_{1}$ is the
``logical projector'' and $\overline{\boldsymbol{A}}=\boldsymbol{I}-\boldsymbol{\Pi}_{1}=\boldsymbol{\Pi}_{0}$
its complement. The ``True'' operator corresponds here to the identity
operator $\boldsymbol{I}$. The ``False'' observable corresponds
to the null operator. These four observables form a complete family
of commuting projectors. The extension to more arguments is obtained
by using the Kronecker product $\otimes$ in the same way as for the
composition of quantum systems (for technical details on this operation
see for example \cite{NC00}).

\noindent In the case of $n=2$ arguments we will have an expansion
over $4$ commuting orthogonal rank-$1$ projectors . Some properties
of the Kronecker product on projectors have to be specified: (i) The
Kronecker product of two projectors is also a projector; (ii) If projectors
are rank-$1$ projectors (a single eigenvalue is equal to~$1$, all
the others are $0$) then their Kronecker product is also a rank-$1$
projector. Using these properties, the $4$ commuting orthogonal rank
-$1$ projectors $\boldsymbol{\Pi}_{00}$, $\boldsymbol{\Pi}_{01}$,
$\boldsymbol{\Pi}_{10}$, and $\boldsymbol{\Pi}_{11}$, spanning the
$4$ dimensional vector space are calculated in a straightforward
way: 
\[
\left\{ \begin{array}{l}
\boldsymbol{\Pi}_{00}\,\,=\,\,(\boldsymbol{I}-\boldsymbol{\Pi})\otimes(\boldsymbol{I}-\boldsymbol{\Pi})\,,\quad\boldsymbol{\Pi}_{01}\,\,=\,\,(\boldsymbol{I}-\boldsymbol{\Pi})\otimes\boldsymbol{\Pi}\,,\\
\boldsymbol{\Pi}_{10}\,\,=\,\,\boldsymbol{\Pi}\otimes(\boldsymbol{I}-\boldsymbol{\Pi})\,,\qquad\qquad\boldsymbol{\Pi}_{11}\,\,=\,\,\boldsymbol{\Pi}\otimes\boldsymbol{\Pi}\,.
\end{array}\right.
\]
So one can write the logical observable for $n=2$ arguments: 
\begin{equation}
\boldsymbol{F}=f(0,0)\:\boldsymbol{\Pi}_{00}+f(0,1)\:\boldsymbol{\Pi}_{01}+f(1,0)\:\boldsymbol{\Pi}_{10}+f(1,1)\:\boldsymbol{\Pi}_{11}\,.\label{eq:OperatDecomp2}
\end{equation}
In an explicit way: 
\[
\boldsymbol{F}\,\,=\,\,\begin{pmatrix}f(0,\,0) & 0 & 0 & 0\\
0 & f(0,\,1) & 0 & 0\\
0 & 0 & f(1,\,0) & 0\\
0 & 0 & 0 & f(1,\,1)
\end{pmatrix}\,.
\]
Equation (\ref{eq:OperatDecomp2}) represents a spectral decomposition
with the eigenvalues being the truth values, in this case we will
have a family of $16$ possible different observables. All these observables
are pairwise commuting projectors and in general their product (matrix
product) is not equal to zero. This last point is essential in the
model, because not only mutually exclusive projectors are representative
for a logical system, the complete family of projectors must be used.
For example the observables for conjunction, $\mathrm{AND}$, and
disjunction, $\mathrm{OR}$, which have in common the truth value,
$(1,1)$, for the input combination $(\mathrm{True}\equiv1\,,\,\mathrm{True}\equiv1)$,
have their matrix product different from zero.

\noindent This method can be extended to whatever number of arguments
$n$ using the ``seed'' projector $\boldsymbol{\Pi}$, its complement
$(\boldsymbol{I}-\boldsymbol{\Pi})$ and by applying the Kronecker
product. So given the number of input arguments $n$ and knowing the
truth table of the logical connective one directly obtains the corresponding
binary Eigenlogic observable.

\noindent Now let's develop the case for $n=2$ arguments: one can
express the connectives corresponding to a ``logical projector''
according to the composition rule, thus obtaining two commuting projector
observables: 
\begin{equation}
\boldsymbol{A}\,\,=\,\,\boldsymbol{\Pi}\otimes\boldsymbol{I}\,,\qquad\boldsymbol{B}\,\,=\,\,\boldsymbol{I}\otimes\boldsymbol{\Pi}\,,\qquad\boldsymbol{A}\cdot\boldsymbol{B}\,\,=\,\,\boldsymbol{\Pi}\otimes\boldsymbol{\Pi}\label{eq:logic-proj-2}
\end{equation}
the conjunction, $\mathrm{AND}$, observable becomes simply the product
of these two logical projectors $\boldsymbol{A}\cdot\boldsymbol{B}$.
The disjunction, $\mathrm{OR},$ and exclusive disjunction, $\mathrm{XOR}$,
observables are shown on Table~1, where the algebraic expansions
for Boolean connectives explicitly derived in \cite{To15} are used.
Negation (complementation) is obtained by subtracting from the identity
operator for projective logical observables and by multiplying by
$\mathrm{-1}$ for isometric logical observables (see hereafter).
Useful transformations are obtained by De Morgan's theorem (for general
theorems in logic see for example Knuth~\cite{Kn09}), for the negative
conjunction, $\mathrm{NAND}$ one has the identity $\overline{A\wedge B}=\overline{A}\vee\overline{B}$
in the same way one can obtain $\mathrm{NOR}$ with the identity $\overline{A\vee B}=\overline{A}\wedge\overline{B}$.
Implication observables are also shown on Table~1.\\
\\
\\

\begin{center}
\begin{tabular}{|c|c|c|c|}
\hline 
connective for  & truth table  & $\{0,1\}$ projective  & $\{+1,-1\}$ isometric \tabularnewline
Boolean  & $\{\mathrm{F},\mathrm{T}\}:$  & logical  & logical \tabularnewline
$A,B$  & $\{0,1\}$ ; $\{+1,-1\}$  & observable  & observable \tabularnewline
\hline 
\hline 
False F  & F F F F  & $\boldsymbol{\mathit{0}}$  & $\boldsymbol{+I}$ \tabularnewline
\hline 
NOR ; $\overline{A\vee B}$  & F F F T  & $\boldsymbol{I}-\boldsymbol{A}-\boldsymbol{B}+\boldsymbol{A}\cdot\boldsymbol{B}$  & $\frac{1}{2}(\boldsymbol{+I}-\boldsymbol{U}-\boldsymbol{V}-\boldsymbol{U}\cdot\boldsymbol{V})$ \tabularnewline
\hline 
$A\,\nLeftarrow\,B$  & F F T F  & $\boldsymbol{B}-\boldsymbol{A}\cdot\boldsymbol{B}$  & $\frac{1}{2}(\boldsymbol{+I}-\boldsymbol{U}+\boldsymbol{V}\text{+}\boldsymbol{U}\cdot\boldsymbol{V})$ \tabularnewline
\hline 
$\overline{A}$  & F F T T  & $\boldsymbol{I}-\boldsymbol{A}$  & $-\boldsymbol{U}$ \tabularnewline
\hline 
$A\nRightarrow B$  & F T F F  & $\boldsymbol{A}-\boldsymbol{A}\cdot\boldsymbol{B}$  & $\frac{1}{2}(\boldsymbol{+I}+\boldsymbol{U}-\boldsymbol{V}+\boldsymbol{U}\cdot\boldsymbol{V})$ \tabularnewline
\hline 
$\overline{B}$  & F T F T  & $\boldsymbol{I}-\boldsymbol{B}$  & $-\boldsymbol{V}$ \tabularnewline
\hline 
XOR ; $A\oplus B$  & F T T F  & $\boldsymbol{A}+\boldsymbol{B}-2\boldsymbol{A}\cdot\boldsymbol{B}$  & $\boldsymbol{U}\cdot\boldsymbol{V}=\boldsymbol{Z}\otimes\boldsymbol{Z}$ \tabularnewline
\hline 
NAND ; $\overline{A\wedge B}$  & F T T T  & $\boldsymbol{I}-\boldsymbol{A}\cdot\boldsymbol{B}$  & $\frac{1}{2}(\boldsymbol{-I}-\boldsymbol{U}-\boldsymbol{V}+\boldsymbol{U}\cdot\boldsymbol{V})$ \tabularnewline
\hline 
AND ; $A\wedge{B}$  & T F F F  & $\boldsymbol{A}\cdot\boldsymbol{B}=\boldsymbol{\Pi}\otimes\boldsymbol{\Pi}$  & $\frac{1}{2}(+\boldsymbol{I}+\boldsymbol{U}+\boldsymbol{V}-\boldsymbol{U}\cdot\boldsymbol{V})$ \tabularnewline
\hline 
$A\equiv B$  & T F F T  & $\boldsymbol{I}-\boldsymbol{A}-\boldsymbol{B}+2\boldsymbol{A}\cdot\boldsymbol{B}$  & $\boldsymbol{-U}\cdot\boldsymbol{V}$ \tabularnewline
\hline 
$B$  & T F T F  & $\boldsymbol{B}=\boldsymbol{I}\otimes\boldsymbol{\Pi}$  & $\boldsymbol{V}=\boldsymbol{I}\otimes\boldsymbol{Z}$ \tabularnewline
\hline 
$A\Rightarrow B$  & T F T T  & $\boldsymbol{I}-\boldsymbol{A}+\boldsymbol{A}\cdot\boldsymbol{B}$  & $\frac{1}{2}(\boldsymbol{-I}-\boldsymbol{U}+\boldsymbol{V}-\boldsymbol{U}\cdot\boldsymbol{V})$ \tabularnewline
\hline 
$A$  & T T F F  & $\boldsymbol{A}=\boldsymbol{\Pi}\otimes\boldsymbol{I}$  & $\boldsymbol{U}=\boldsymbol{Z}\otimes\boldsymbol{I}$ \tabularnewline
\hline 
$A\Leftarrow B$  & T T F T  & $\boldsymbol{I}-\boldsymbol{B}+\boldsymbol{A}\cdot\boldsymbol{B}$  & $\frac{1}{2}(\boldsymbol{-I}+\boldsymbol{U}-\boldsymbol{V}-\boldsymbol{U}\cdot\boldsymbol{V})$ \tabularnewline
\hline 
OR ; $A\vee B$  & T T T F  & $\boldsymbol{A}+\boldsymbol{B}-\boldsymbol{A}\cdot\boldsymbol{B}$  & $\frac{1}{2}(\boldsymbol{-I}+\boldsymbol{U}+\boldsymbol{V}+\boldsymbol{U}\cdot\boldsymbol{V})$ \tabularnewline
\hline 
True T  & T T T T  & $\boldsymbol{I}$  & $-\boldsymbol{I}$ \tabularnewline
\hline 
\end{tabular}
\par\end{center}

\begin{center}
\textbf{Table 1.} The sixteen two-argument two-valued logical connectives
and the respective Eigenlogic observables for eigenvalues $\{0,1\}$
and $\{+1,-1\}$. 
\par\end{center}

\subsection{Isometric reversible two-valued logical observables}

There is a linear bijection (isomorphism) from the projector logical
observables $\boldsymbol{F}$ towards reversible observables $\boldsymbol{G}$:
\[
\boldsymbol{G}=\boldsymbol{I}-2\boldsymbol{F}\,.
\]
The two families of observables commute and have the same system of
eigenvectors. Practically to obtain $\boldsymbol{G}$ from $\boldsymbol{F}$
one just has to substitute the eigenvalue $0$ with $+1$ and $1$
with $-1$. The observables $\boldsymbol{G}$ are ``isometries'':
unitary reflection operators. From projector $\boldsymbol{\Pi}$ in
equation (\ref{eq:logic-proj-2}) one obtains the observable $\boldsymbol{Z}$:
\[
\boldsymbol{Z}=\boldsymbol{I}-2\boldsymbol{\Pi}=\left(\begin{array}{cc}
+1 & 0\\
0 & -1
\end{array}\right)=\sigma_{z}
\]
which is actually one of the Pauli matrices $\sigma_{z}$ and corresponds
in quantum mechanics, to the $z$ component of a spin $\nicefrac{1}{2}$
observable $\mathbf{S}_{z}=(\nicefrac{\hbar}{2})\,\sigma_{z}$ where
$\hbar$ is the reduced Planck's constant. In the field of quantum
information this operator is also named the ``Pauli-$Z$'' gate
or ``phase-$\pi$'' gate \cite{NC00}. Here, $\boldsymbol{U}=\boldsymbol{Z}$
designates the ``logical projector'' connective and $\,\overline{\boldsymbol{U}}=-\boldsymbol{Z}$
its complement (negation), \textit{nota bene} in this case the connective
``logical projector'' is not a projection operator, in order to
avoid ambiguity it is often named \cite{MO08} ``dictator''.

For $n=2$ arguments one can then write directly the expression for
a logical isometric observable by using its spectral decomposition.
The logical ``dictators'' $\boldsymbol{U}$ and $\boldsymbol{V}$
become: 
\[
\boldsymbol{U}\,\,=\,\,\boldsymbol{Z}\otimes\boldsymbol{I}\,,\qquad\boldsymbol{V}\,\,=\,\,\boldsymbol{I}\otimes\boldsymbol{Z}\,,\qquad\boldsymbol{U}\cdot\boldsymbol{V}\,\,=\,\,\boldsymbol{Z}\otimes\boldsymbol{Z}\,.
\]
The exclusive disjunction $XOR$ observable is here simply given by
the product of the dictators: $\boldsymbol{U}\cdot\boldsymbol{V}$.
Negation is obtained by multiplying by the number $-1$. From table~1
one sees that there are more complicated relations, for example the
conjunction, $AND$, observable is: 
\[
\frac{1}{2}(\boldsymbol{I}+\boldsymbol{U}+\boldsymbol{V}-\boldsymbol{U}\cdot\boldsymbol{V})=\left(\begin{array}{cccc}
+1 & 0 & 0 & 0\\
0 & +1 & 0 & 0\\
0 & 0 & +1 & 0\\
0 & 0 & 0 & -1
\end{array}\right)\,\,=\,\,\boldsymbol{C}^{Z}\,.
\]
Those familiar with the domain of quantum information can easily recognize
the reversible logical gate ``control-$Z$'' or simply named $C^{Z}$
\cite{NC00}.

\section{From deterministic logic to fuzzy logic}

Fuzzy logic deals with truth values that may be any number between
0 and 1, here the truth of a proposition may range between completely
true and completely false. It is generally considered that probability
theory and fuzzy logic are related to different forms of uncertainty,
the first is concerned with how probable it is that a variable belongs
to a given set and the second one uses the concept of fuzzy set membership,
intended as the degree of membership. This was the first motivation
of fuzzy logic \cite{key-5}. But this distinction when considering
the quantum probabilistic Born rule is not so strict from a formal
point of view. We will start the discussion by giving the interpretation
of a vector state in Eigenlogic.

In the preceding sections we considered operations on the eigenspace
of a logical observable family. For example for $n=2$ arguments a
complete family of $16$ commuting logical observables represents
all possible logical connectives and becomes ``interpretable'' \cite{To15}
when applied to one of the four possible canonical eigenvectors of
the family. These vectors, corresponding to all the possible atomic
input propositional cases, are represented by the vectors $|00>$,
$|01>$, $|10>$ and $|11>$ forming a complete orthonormal basis.
When applying a logical observable on one of these vectors the resulting
eigenvalue will correspond to the truth value for the considered input.

\noindent Now what happens when the state-vector is not one of the
eigenvectors of the logical system? In quantum mechanics, where vectors
operate in Hilbert space, one can always express a state-vector as
a decomposition on a complete orthonormal basis. In particular we
can express it over the canonical eigenbasis of the logical observable
family. For two-arguments this vector can be written as: 
\[
|\Psi>\,\,=\,\,C_{00}\,|00>\,+\,C_{01}\,|01>\,+\,C_{10}\,|10>\,+\,C_{11}\,|11>\,.
\]
We can interpret this in the following way: when only one of the coefficients
is non-zero (in this case its absolute value must take the value $1$)
then we are back in the preceding situation of a determinate input
atomic propositional case. But when more than one coefficient is non-zero
we are in a ``mixed'' or ``fuzzy'' propositional case. Such a
state could also possibly be interpreted as a quantum superposition
of atomic propositional cases.

\noindent We can then calculate the ``mean value'' of a logical
observable. In particular the logical projector observables $\boldsymbol{F}$
will give a ``fuzzy measure'' of the logical proposition in the
form of the ``fuzzy membership function'' $\mu$. Let's show this
on some examples: in the case of one argument one can express an arbitrary
$2$-dimensional quantum state as: $|\varphi>=\sin\alpha\,|0>+e^{i\beta}\,\cos\alpha\,|1>$
where the ``angles'' $\alpha$ and $\beta$ are real numbers. The
quantum mean value of the ``logical projector'' observable $\boldsymbol{A}=\boldsymbol{\Pi}$
can then be calculated using the Born rule: 
\[
\mu(a)\,\,=\,\,<\varphi|\boldsymbol{\Pi}|\varphi>\,\,=\,\,\cos\alpha\,\,\mathrm{e^{-i\beta}}<1|1>\,<1|\cos\alpha\mathrm{e^{i\beta}}|1>\,\,=\,\,\cos^{2}\alpha\,;
\]
in the same way one can calculate the complement 
\[
\mu(\overline{a})\,\,=\,\,<\varphi|\boldsymbol{I}-\boldsymbol{\Pi}|\varphi>\,\,=\,\,\sin^{2}\alpha\,\,=\,\,1-\mu(a)\,.
\]
This verifies one of the requirements of fuzzy logic for the complement
(negation) of a fuzzy set.

\noindent According to standard notations for spin $\nicefrac{1}{2}$
quantum states, or qubits, on the Bloch sphere \cite{NC00} we use
the transformation $\alpha=\nicefrac{\left(\pi-\theta\right)}{2}$
and $\beta=\varphi$. A quantum compound state can be built by taking
the tensor product of two elementary states: $|\psi>=|\varphi_{p}>\otimes|\varphi_{q}>$,
where $|\varphi_{p}\!>=\cos\frac{\theta_{p}}{2}|0>+e^{i\varphi_{p}}\sin\frac{\theta_{p}}{2}|1>$
(for $|\varphi_{q}>$ we have a similar expression). Now $\sin^{2}\frac{\theta_{p}}{2}=p$
and $\sin^{2}\frac{\theta_{q}}{2}=q$ represent the probabilities
of being in the ``True'' state $|1>$ for spins $\nicefrac{1}{2}$
oriented along two different axes $\theta_{p}$ and $\theta_{q}$
.

\noindent One can calculate the fuzzy membership function of the corresponding
``logical projector'' for the two-argument case using equation (\ref{eq:logic-proj-2}).
\[
\mu(a)=<\psi|\boldsymbol{\Pi}\otimes\boldsymbol{I}|\psi>=p(1-q)+p\cdot q=p\,,\quad\mu(b)=<\psi|\boldsymbol{I}\otimes\boldsymbol{\Pi}|\psi>=q\,.
\]
This shows that the mean values correspond to the respective probabilities.
Now let's ``measure'' for example the conjunction and the disjunction,
using the observables in table~1, this gives: 
\[
\left\{ \begin{array}{l}
\mu(a\mathrm{\wedge}b)\,\,=\,\,<\psi|\boldsymbol{\Pi}\otimes\boldsymbol{\Pi}|\psi>\,\,=\,\,p\cdot q\,\,=\,\,\mu(a)\cdot\mu(b)\,,\\
\mu(a\mathrm{\vee}b)\,\,=\,\,p+q-p\cdot q\,\,=\,\,\mu(a)+\mu(b)-\mu(a)\cdot\mu(b)\,.
\end{array}\right.
\]
Similar results for conjunction and disjunction have been outlined
recently, also using projector operators, when considering concept
combinations \cite{ASV15} for quantum-like experiments in the domain
of quantum cognition.

\noindent What happens when the state-vector cannot be put in the
form of a tensor product, that is when it corresponds to an entangled
state? The problem is outside the scope of this paper but an interesting
result can be shown: the mean value of whatever logical observable
of the type $\boldsymbol{F}$ on an arbitrary quantum state $|\Psi>$
will always verify the inequality: 
\[
<\Psi|\boldsymbol{F}|\Psi>\,\,=\,\,\mathrm{Tr}\,(\rho_{\Psi}\cdot\boldsymbol{F})\,\leq\,1\,,\qquad{\rm with}\quad\rho_{\Psi}\,\,\equiv\,\,|\Psi>\,<\Psi|\,,
\]
and can thus be interpreted as a probability measure.

\section{From two-valued to multi-valued logic}

Multi-valued logic requires a different algebraic structure than an
ordinary binary-valued one. Many properties of binary logic do not
support set of values that do not have cardinality $2^{n}$. Multi-valued
logic is often used for the development of logical systems that are
more expressive than Boolean systems for reasoning \cite{MT08}. Particularly
three and four valued systems, have been of interest with applications
to digital circuits and computer science.

The total number of possible logical connectives for an $m$-valued
$n$-argument system is the combinatorial number $m^{m^{n}}$, so
in particular for a binary $2$-valued $2$-argument system, as shown
above, the number of connectives will be $2^{2^{2}}=16$, the complete
list indicated on table~1. For a binary three-argument system, the
number increases to $2^{2^{3}}=256$. For a $3$-valued $1$-argument
system the number of connectives will be $3^{3^{1}}=27$ and for a
$3$-valued $2$-argument system: $3^{3^{2}}=19683$. So it is clear
that by increasing the values from two to three the possibilities
of new connectives becomes intractable for a complete description
of a logical system, but some special connectives play important roles
and will be illustrated hereafter. We will proceed by showing the
general algebraic method.

\subsection{Interpolation with finite elements}

The finite element method (see for example \cite{Zi71}) allows one
to interpolate a function, \textit{id est} to make explicit the values
$f(x)$ from the given values of specific numbers, the (so-called)
degrees of freedom.

Let's consider the following simple example: given the values $f(\text{+}1)$,
$f(0)$ and $f(-1)\,$ of a function $f$ at the particular points
$\,x=+1,\,0,\,-1$, and using the appropriate Dirac linear forms,
we can write: $<\delta_{+1}\,,\,f>=f(+1)$ , $<\delta_{0}\,,\,f>=f(0)$
and $<\delta_{-1}\,,\,f>=f(-1)$, where $\Sigma\equiv\{\,\delta_{+1}\,,\,\delta_{0}\,,\,\delta_{-1}\,\}$
is called the set of degrees of freedom. This linear structure shows
that it is natural to consider a three-dimensional space. The so-called
``basis function'' $\varphi_{i}$ associated to the set of degrees
of freedom $\Sigma$ and to the polynomial space solves this problem.
Using the degrees of freedom, and second-degree polynomials, we obtain
the three basis functions 
\begin{equation}
\varphi_{+1}(x)=\frac{1}{2}\,x\,(x+1)\,,\quad\varphi_{0}(x)=1-x^{2}\,,\quad\varphi_{-1}(x)=\frac{1}{2}\,x\,(x-1)\,.\label{base-P2}
\end{equation}
So in general, an arbitrary function $f$ can be written: 
\begin{equation}
f(x)=\sum_{i=+1,0,-1}f(i)\,\,\,\varphi_{i}(x)\,,\qquad\qquad\sum_{i=+1,0,-1}\,\varphi_{i}(x)\,\equiv\,1\,\label{decomp-f}
\end{equation}
where the completeness of the basis functions is verified by their
sum being $1$.

\subsection{Formalization of three-valued Eigenlogic}

We use an operator system which is equivalent to the one of orbital
angular momentum $\ell=1$. In general angular momentum is characterized
by two quantum numbers: $j$ the angular momentum number and $m_{j}$
the magnetic momentum number. Both these numbers must be integer or
half integer. The rules are: $j\geq0$, and attached to this value
we have the condition: $-j\leq m_{j}\leq j$. The value $j=0$ is
possible and gives a single value $m_{j}=0$ the next is $j=s=\nicefrac{1}{2}$
giving two values $m_{s}=\pm\nicefrac{1}{2}$ corresponding to the
two-valued spin system. The value $j=1$ gives three possible values
$m_{j}=\{+1,0,-1$\} and so on. We consider for $j=\ell=1$ the $z$-component
orbital angular momentum observable \cite{Sc49} 
\begin{equation}
\boldsymbol{L}_{z}=\hbar\boldsymbol{\varLambda}=\hbar\left(\begin{array}{ccc}
+1 & 0 & 0\\
0 & 0 & 0\\
0 & 0 & -1
\end{array}\right)\,.\label{eq:Lz}
\end{equation}
In the above matrix the three eigenvalues $\{+1,0,-1\}$ will be considered
as the logical values. A convention for these values, extending binary
logic, is the following:

\begin{center}
\quad{}${\bf \mathrm{False}}:\,\mathrm{F}\equiv+1\,,\,\,\mathrm{{\rm Neutral:}}\,\mathrm{N}\equiv0\,,\,\,{\rm True:}\,\mathrm{T}\equiv-1\,.$
\par\end{center}

We can now express the three-value logical observables as spectral
decompositions over the rank-$1$ projectors spanning the vector space:
$\boldsymbol{\Pi}_{+1}$, $\boldsymbol{\Pi}_{0}$ and $\boldsymbol{\Pi}_{-1}$.
These operators correspond to the pure state density matrices of the
three eigenstates $|+1>$ , $|0>$ and $|-1>$ of $\boldsymbol{L}_{z}$.
The three projectors can be expressed as a function of the dimensionless
observable $\boldsymbol{\varLambda}$, using directly the expressions
given above in (\ref{base-P2}) where the basis functions $\varphi_{i}$
become the projectors and the symbol $x$ the observable $\boldsymbol{\varLambda}$
given in (\ref{eq:Lz}): 
\begin{equation}
\boldsymbol{\Pi}_{+1}=\frac{1}{2}\boldsymbol{\varLambda}\left(\boldsymbol{\varLambda}+\boldsymbol{I}\right)\qquad\boldsymbol{\Pi}_{0}=\boldsymbol{I}-\boldsymbol{\varLambda}^{2}\qquad\boldsymbol{\Pi}_{-1}=\frac{1}{2}\boldsymbol{\varLambda}\left(\boldsymbol{\varLambda}-\boldsymbol{I}\right)\label{eq:Project-3v}
\end{equation}
Then every one-argument ``local projector'' $\boldsymbol{F}(\boldsymbol{\varLambda})$
can be obtained using the relation (\ref{decomp-f}).

\subsection{Three-valued, two-argument examples: Min, Max}

When considering a $2$-argument $3$-valued system we find the expansion
by using the Kronecker product in the same way as for the binary system
in equation (\ref{eq:OperatDecomp2}): 
\begin{equation}
\boldsymbol{F}\,\,=\,\,\sum_{i,\,j\,\,=+1,\,0,\,-1}f_{ij}\,\,\boldsymbol{\Pi}_{i}\otimes\,\boldsymbol{\Pi}_{j}\,,\qquad f_{ij}\in\{+1,\,0,\,-1\}\,.\label{interpol-2}
\end{equation}
these observables are now $9\times9$ matrices. We can define the
two argument ``dictators'', $\boldsymbol{U}$ and $\boldsymbol{V}$,
simply by the rule of composition, this leads to: 
\begin{equation}
\boldsymbol{U}=\boldsymbol{\varLambda}\otimes\boldsymbol{I}\qquad\qquad\boldsymbol{V}=\boldsymbol{I}\otimes\boldsymbol{\varLambda}\qquad\qquad\boldsymbol{U}\cdot\boldsymbol{V}=\boldsymbol{\varLambda}\otimes\boldsymbol{\varLambda}\,.\label{eq:Dictat-3v 2a}
\end{equation}
In trivalent logic (see \textit{e.g.} \cite{MT08}) popular connectives
are $\mathrm{Min}$ and $\mathrm{Max}$, defined in the maps on table~2.

Here the connectives $\mathrm{Min}$ and $\mathrm{Max}$ are symmetric,
they are equivalent for a complete inversion of signs on inputs and
outputs. Using the relations (\ref{eq:Project-3v}), (\ref{interpol-2})
and (\ref{eq:Dictat-3v 2a}) in conjunction with reduction rules we
obtain the following observables:

\begin{center}
\begin{equation}
\left\{ \begin{array}{l}
{\displaystyle {\rm Min}(\boldsymbol{U},\boldsymbol{V})\,=\,\frac{1}{2}\,\big(\boldsymbol{U}+\boldsymbol{V}+\boldsymbol{U}^{2}+\boldsymbol{V}^{2}-\boldsymbol{U}\cdot\boldsymbol{V}-\boldsymbol{U}^{2}\cdot\boldsymbol{V}^{2}\big)}\\
Max(\boldsymbol{U},\boldsymbol{V})\,\,=\,\,\frac{1}{2}\,\big(\boldsymbol{U}+\boldsymbol{V}-\boldsymbol{U}^{2}-\boldsymbol{V}^{2}+\boldsymbol{U}\cdot\boldsymbol{V}+\boldsymbol{U}^{2}\cdot\boldsymbol{V}^{2}\big)
\end{array}\right.\label{minmax}
\end{equation}

\par\end{center}

\noindent 

\begin{center}
\begin{tabular}{|c|c|c|c|}
\hline 
Min \,\, $U\,\,\backslash\!\!\backslash\,\,V$  & \quad{}F ~  & \quad{}N ~  & \quad{}T ~ \tabularnewline
\hline 
\hline 
F $\equiv$ +1  & +1  & +1  & +1 \tabularnewline
\hline 
N $\equiv$ 0  & +1  & 0  & 0 \tabularnewline
\hline 
T $\equiv$ $-$1  & +1  & 0  & $-$1 \tabularnewline
\hline 
\end{tabular}\qquad{}~\ %
\begin{tabular}{|c|c|c|c|}
\hline 
Max \,\, $U\,\,\backslash\!\!\backslash\,\,V$  & \quad{}F ~  & \quad{}N ~  & \quad{}T ~ \tabularnewline
\hline 
\hline 
F $\equiv$ +1  & +1  & 0  & $-$1 \tabularnewline
\hline 
N $\equiv$ 0  & 0  & 0  & $-$1 \tabularnewline
\hline 
T $\equiv$ $-$1  & $-$ 1  & $-$1  & $-$1 \tabularnewline
\hline 
\end{tabular}
\par\end{center}

\noindent \begin{center}
\textbf{Table 2.} The Min and Max maps for a three-valued two-argument
logic.
\par\end{center}

\pagebreak{}

The proof of the relations (\ref{minmax}) is a direct consequence
of relations (\ref{interpol-2}) and (\ref{base-P2}). We have on
one hand:\\

\noindent ${\displaystyle {\rm Min}\,(U,V)\,=\,\varphi_{1}(U)\,\otimes\,\varphi_{1}(V)\,+\,\varphi_{1}(U)\,\otimes\,\varphi_{0}(V)\,+\,\varphi_{1}(U)\,\otimes\,\varphi_{-1}(V)}$

\noindent ${\displaystyle \qquad\qquad\qquad\,+\,\varphi_{0}(U)\,\otimes\,\varphi_{1}(V)\,+\,\varphi_{-1}(U)\,\otimes\,\varphi_{1}(V)\,-\,\varphi_{-1}(U)\,\otimes\,\varphi_{-1}(V)}$

\noindent ${\displaystyle =\varphi_{1}(U)+\varphi_{1}(V)\,-\,\varphi_{1}(U)\,\otimes\,\varphi_{1}(V)\,-\,\varphi_{-1}(U)\,\otimes\,\varphi_{-1}(V)}$
\hfill{}due to (\ref{decomp-f})

$=\frac{1}{2}\,U\,(U+I)\,+\,\frac{1}{2}\,V\,(V+I)\,-\,\frac{1}{4}\,U\,(U+I)\,V\,(V+I)\,-\,\frac{1}{4}\,U\,(U-I)\,V\,(V-I)$

\noindent $=\frac{1}{2}\,\big(U^{2}\,+\,U\,+\,V^{2}\,+\,V\,-\,U^{2}V^{2}\,-\,UV\big)$\\

\noindent and the first relation of (\ref{minmax}) is proven. On
the other hand, we have\\

\noindent ${\displaystyle {\rm Max}\,(U,V)\,=\,\varphi_{1}(U)\,\otimes\,\varphi_{1}(V)\,-\,\varphi_{1}(U)\,\otimes\,\varphi_{-1}(V)\,-\,\varphi_{0}(U)\,\otimes\,\varphi_{-1}(V)}$

\noindent ${\displaystyle \qquad\qquad\qquad\,-\,\varphi_{-1}(U)\,\otimes\,\varphi_{-1}(V)\,-\,\varphi_{-1}(U)\,\otimes\,\varphi_{1}(V)\,-\,\varphi_{-1}(U)\,\otimes\,\varphi_{0}(V)}$

\noindent ${\displaystyle =\varphi_{1}(U)\,\otimes\,\varphi_{1}(V)\,-\,\varphi_{-1}(U)\,-\,\varphi_{-1}(V)\,+\,\varphi_{-1}(U)\,\otimes\,\varphi_{-1}(V)}$
\hfill{}due to (\ref{decomp-f}) 

\noindent $=\frac{1}{4}\,U\,(U+I)\,V\,(V+I)\,-\,\frac{1}{2}\,U\,(U-I)\,-\,\frac{1}{2}\,V\,(V-I)+\frac{1}{4}\,U\,(U-I)\,V\,(V-I)$

$=\frac{1}{2}\,\big(U^{2}V^{2}+UV\,-\,U^{2}\,-\,V^{2}\,+\,U\,+\,V\big)$

\noindent and the second relation of (\ref{minmax}) is proven. \hfill{}$\square$ 

\noindent The proof presented above exploits the properties of the
Kronecker product and reduction rules due to the completeness of the
finite projection space. Reduction of logical expressions is an important
topic in logic. In binary logic it is formalized by using Karnaugh
maps which represent canonical SOP (Sum Of Products) disjunctive normal
forms \cite{Kn09}.

\noindent Binary logic is ``included'' in ternary logic, we want
to verify this by eliminating the ``neutral'' state, $\mathrm{N}\equiv0$,
and considering only the two logical values $\{+1,-1\}$. In this
case we have: $\,\boldsymbol{U}^{2}=\boldsymbol{V}^{2}=\boldsymbol{I}$
and so (\ref{minmax}) reduces to: 
\[
\left\{ \begin{array}{l}
{\rm Min}\,(\boldsymbol{U},\boldsymbol{V})\,=\,\frac{1}{2}\,\big(\boldsymbol{I}+\boldsymbol{U}+\boldsymbol{V}-\boldsymbol{U}\cdot\boldsymbol{V}\big)\,,\\
\,\,{\rm Max}\,(\boldsymbol{U},\boldsymbol{V})\,\,=\,\,\frac{1}{2}\,\big(-\boldsymbol{I}+\boldsymbol{U}+\boldsymbol{V}+\boldsymbol{U}\cdot\boldsymbol{V}\big)
\end{array}\right.
\]
considering that for binary logic the $\mathrm{Min}$ connective becomes
the conjunction, $\mathrm{AND}$, and the $\mathrm{Max}$ connective
the disjunction, $\mathrm{OR}$, we find the previous results given
on table~1 for binary $\{+1,-1\}$ observables.

\section{Discussion and Conclusion}

We have presented an operational formalism named ``Eigenlogic''
using observables in Hilbert space. The original feature being that
the eigenvalues of a logical observable represent the truth values
of the corresponding logical connective, the associated eigenvectors
corresponding to one of the fixed combination of the inputs (atomic
propositions). This approach differs from other geometric formalizations
of logic (for references and discussion see \cite{To15}). Here the
outcome of a ``measurement'' or ``observation'' on a logical observable
will give the truth value of the associated logical proposition, and
becomes ``interpretable'' when applied to the eigenspace leading
to a natural analogy with the measurement postulate in quantum mechanics.
One of the referees proposed the following diagram to summarize the
point of view presented in this contribution:\\

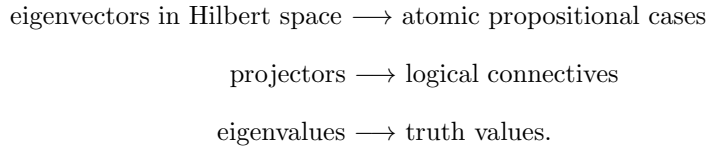
\begin{center}
eigenvectors in Hilbert space $\longrightarrow$ atomic propositional
cases
\par\end{center}

\begin{center}
~~~\qquad{}\qquad{}projectors $\longrightarrow$ logical connectives
\par\end{center}

\begin{center}
\qquad{}eigenvalues $\longrightarrow$ truth values.\\

\par\end{center}

At first sight this method could be viewed as ``classical'' because
exactly the same results are obtained in Eigenlogic as in ordinary
propositional logic. This is in itself an important result demonstrating
a new method in logic based on linear algebra, the method being also
developed in multivalued logic. But when considering vector states,
\textit{id est} input propositions, that are not eigenvectors, the
measurement outcomes are governed by the quantum Born rule, and interpretable
results are then given by the mean values. This fact led us to apply
the method to Fuzzy logic.

\noindent Another important point is the general algebraic method,
based on classical interpolation framework suggested by the finite-element
method. Our method can be employed for whatever $m$-valued $n$-argument
logical system and in each case the corresponding logical observables
can be defined. Some observables can be formally compared with angular
momentum observables in quantum mechanics. Because of the exponential
increase of complexity, an analytical formulation is only tractable
for a low number of logical values and arguments. We treated the two-argument
binary case completely and the three-valued case using the logical
observables $\mathrm{Min}$ and $\mathrm{Max}$. An algorithmic approach
for logical connectives with a large number of arguments could be
interesting to develop using Eigenlogic observables in high-dimensional
vector spaces. But because the space grows in dimension very quickly,
it may not be particularly useful for practical implementation without
logical reduction. It would be interesting to develop specific algebraic
reduction methods for logical observables inspired from actual research
in the field. For a good synthesis of the state of the art, see \textit{e.g.}
\cite{YS08}.

\noindent Eigenlogic could create a new perspective in the field of
quantum computation because several of the observables turn out to
be well-known quantum gates. Here we represent them as diagonal matrices,
\textit{id est} in their eigenbasis, other ``normal'' forms being
easily recovered by unitary transformations. It would be interesting
to operate quantum gates in our framework. Many-valued logic is being
investigated in quantum computation for example with ternary-logic
quantum gates using ``qutrits''. Our formulation of multivalued
logical observables could be used for the design of new quantum gates.

\noindent Dynamical evolution of the logical system could be included
in the model by identifying the appropriate Hamiltonian operators.
Standard procedures for expressing interaction Hamiltonians as a function
of angular momentum observables could be used \cite{Sc49}.

\noindent More generally we think that this view of logic could add
some insight on more fundamental issues. Boolean functions are nowadays
considered as a ``toolbox'' for resolving many problems in theoretical
computer science, information theory and even fundamental mathematics.
In the same way Eigenlogic can be considered as a new ``toolbox''
and could be of interest for the ``Quantum Interaction'' community
where quantum-like approaches in human and social sciences need to
be founded on a logical basis.

\section*{Acknowledgments}

\noindent The authors thank both referees for their precise and constructive
remarks and suggestions. Some of them have been included in the present
version of this contribution.\\
\\
\\
\\


\begin{thebibliography}{10}
\bibitem{BvN1936} Birkhoff, G., von Neumann J.: The Logic of Quantum
Mechanics. The Annals of Mathematics, 2nd Ser., 37 (4), 823-843 (1936)

\bibitem{Yi16} Ying, M.S. Foundations of Quantum Programming. Morgan
Kaufmann (2016) 

\bibitem{BTMD14} Barros, J., Toffano, Z., Meguebli, Y., Doan, B.L.:
Contextual Query Using Bell Tests. Springer Lecture Notes in Computer
Science, LNCS 8369, 110-121 (2014)

\bibitem{BB12} Busemeyer, J.R., Bruza, P.D.: Quantum models of cognition
and decision. Cambridge University Press (2012)

\bibitem{To15} Toffano, Z.: Eigenlogic in the spirit of George Boole.
ArXiv:1512.06632 (2015)

\bibitem{MO08} Montanaro, A., Osborne, T.J.: Quantum Boolean functions.
ArXiv:0810.2435 (2008)

\bibitem{NC00} Nielsen, M.A., Chuang, I.L.: Quantum Computation and
Quantum Information. Cambridge University Press (2000)

\bibitem{Kn09} Knuth, D.E.: The Art of Computer Programming, Volume
4, Fascicle 0: Introduction to Combinatorial Algorithms and Boolean
Functions. Addison-Wesley (2009)

\bibitem{key-5} Zadeh, L.A.: Fuzzy sets. Information and Control,
8 (3), 338-353 (1965)

\bibitem{ASV15} Aerts, D., Sozzo, S., Veloz, T.: Quantum structure
of negation and conjunction in human thought. Front Psychol. 6:1447
(2015)

\bibitem{MT08} Miller, D.M., Thornton, M.A.: Multiple Valued Logic:
Concepts and Representations. Morgan \& Claypool Publishers (2008)

\bibitem{Zi71} Zienkiewicz,O.Z. The Finite Element Method in Engineering
Science. Mc Graw-Hill, New York (1971)

\bibitem{Sc49} Schiff, L.I.: Quantum Mechanics. Mc Graw-Hill, New
York (1949)

\bibitem{YS08} Yanushkevich, S.N., Shmerko, V.P.: Introduction to
Logic Design. CRC Press (2008)\end{thebibliography}
\end{document}